\documentclass[%
 amsmath,amssymb
]{cas-sc}
\usepackage[numbers]{natbib}
\usepackage{graphicx}
\usepackage{dcolumn}
\usepackage{bm}
\usepackage{natbib}
\usepackage{hyperref}

\begin{document}

\title{Landau quantization of a circular Quantum Dot using the BenDaniel-Duke boundary condition}

 \author{Sriram Gopalakrishnan}%
 \address{Indian Institute of Technology Madras,  Chennai, India- 600036}
 \credit{Formal analysis, Software, Visualization, Writing-Original Draft}
 \author{Sayak Biswas}%
 \address{Indian Institute of Science Education and Research, Kolkata, India - 741246
 }
 \credit{Formal analysis, Software, Visualization}
 \author{Shivam Handa}%
 \address{Massachusetts Institute of Technology, Cambridge MA 02139}
\credit{Formal analysis}

\begin{abstract}
  We derive the  energy levels of a circular Quantum  Dot (QD) under a
  transverse  magnetic   field,  incorporating  the   Ben-Daniel  Duke
  boundary  condition  (BDD). The  parameters  in  our model  are  the
  confinement barrier height,  the size of the QD,  the magnetic field
  strength,  and  a  mass  ratio  highlighting  the  effect  of  using
  BDD. Charge  densities, transition  energies, and the  dependence of
  energies on  magnetic field has  been calculated to show  the strong
  influence of  BDD.  We  find that  our numerical  calculations agree
  well with  experimental results on  the GaAs-InGaAs Quantum  Dot and
  can  be used  further.   We also  provide  an insightful  analytical
  approximation  to our  numerical results, which converges well for larger values of size and confinement.
\end{abstract}

\begin{highlights}
\item The energy levels  of a circular  Quantum Dot (QD)  under a
  transverse  magnetic   field,  incorporating  the   Ben-Daniel  Duke
  boundary condition  (BDD) are derived and calculated numerically.

\item Theoretical findings were compared with the previously published experimental results   on the
  GaAs-InGaAs  Quantum  Dot and found to be in agreement.
  
\item An insightful asymptotic approximation is provided, which converges with numerical results for larger values of size and confinement.
\end{highlights}


\maketitle


\section{INTRODUCTION}

Low dimensional quantum systems constitute  an active area of research
with  widespread applications  in technology.   The non-abelian  anyon, for  instance,   is   a   two  dimensional   (2D)
quasiparticle  proposed  as  an   option  for  fault-tolerant  quantum
computation \cite{wen1991non, nayak2008non}.   Quantum Dots  (QDs), also  known as
artificial  atoms \cite{ashoori96},  are  nanostructures that  tightly
confine electrons in quantum wells,  resulting in bound states.  Owing
to their  tunability, QDs have several  applications, including quantum information processing \cite{veldhorst2014,loss98}, and QD based light
emitting devices \cite{anikeeva2009,mashford2013}. A recent experiment
also  probed   the  energy  levels   of  a  QD  in   Bilayer  Graphene \cite{kurzmann2019}. Accurate level schemes of QDs are relevant in the present  context  as  technology inches towards quantum computing.

 Quantum  Dots  are  fabricated  by  forming  heterojunctions  between
 dissimilar  semiconductors \cite{harrison2016quantum}. It may be noted that the core and shell materials of a heterojunction QD can have very similar properties if not for their bandgaps. Depending on the alignment of the valence and conduction band edges across the interface, QDs are classified as Type-I or Type-II \cite{ivanov2007type}. The nature of band edge alignment results in a confinement potential whose profile is usually approximated as a parabola  or  a  finite hard-wall  for  simplicity  of  theoretical modeling.  Although a  parabolic  profile is  less  idealized than  a finite  hard-wall, the  latter allows  us to develop a phenomenology accounting for finite size and barrier height.
 
 Additionally, models must employ Effective Mass Theory (EMT) accurately, so as to account for a spatially varying carrier effective mass created by the confinement potential. Hamiltonians must be modified to maintain hermiticity. In the case of hard-wall confinement, there is a discontinuous change in effective mass across the barrier. The corrected Hamiltonian thus leads to a modified boundary condition on the derivative of the wavefunction, called the BenDaniel Duke boundary condition (BDD) \cite{bdd66}. In this regard, we define a dimensionless mass ratio $\beta=m_i/m_o$, where $m_i$ and $m_o$ are the effective masses of the electron inside and outside the well respectively.


The  goal of  this  paper  is to analytically develop the complete set of spin-degenerate Landau levels of a single-electron, hard-wall confined, circular QD placed in a perpendicular magnetic field using BDD and examining the effect of imposing BDD. There have  been extensive studies on QDs  in the past
three decades.  However, only few  of these models include  and examine
the            effect            of            imposing            BDD
\cite{singh2000role,singh2006revisiting,singh2011approximate,koc15,duja18}.       A
recent  theoretical work  modelled CdSe/CdS  core-shell QDs  using BDD
\cite{nandan2019wavefunction}. The general approach of our theory can be used to understand data obtained in experiments such as Gated Transport Spectroscopy (GTS) and Single Electron Capacitance Spectroscopy (SECS) of Quantum Dots with electrostatic confinement and magnetic fields \cite{ashoori96}.

The system we consider is a single electron trapped in a finite, radially symmetric potential well in 2D, and placed in a perpendicular magnetic field. We have accomodated the possibility of different magnetic fields inside and outside the QD, although we use a uniform magnetic field in numerical calculations. The confinement potential approximates a thin InGaAs quantum disk sandwiched between two layers of GaAs, as experimentally probed by Drexler \textit{et. al.} \cite{drexler94}. A hard-wall confinement model was soon proposed by Peeters \textit{et. al.} \cite{peeters96}, however, they used a two-electron model without BDD to fit a transition gap with experimental data. It is known in practice that the transition gaps of a QD are effectively independent of electron-electron interactions \cite{peeters1990magneto,sikorski1989spectroscopy}. Hence, we propose a single-electron model in conjunction with BDD to find agreement with the same data. We would like to mention that the ground state of our system was studied, including the effect of BDD, by Asnani \textit{et. al.} \cite{asnani2009}. We extend the study using Landau quantization to obtain a complete electronic structure from the Schrodinger equation, and test the effect of imposing BDD on multiple Landau levels created by a homogeneous magnetic field.

QDs are also of interest from a fundamental physics point of view, particularly in understanding non-local phenomena such as the Aharanov-Bohm effect, where charges can be influenced by electromagnetic potentials even in the absence of electromagnetic fields \cite{ahar59,yeyati1995aharonov,yacoby1995coherence}. In the following analysis, although we consider an inhomogeneous magnetic field only to keep the formalism general, there is an interesting phenomenon of magnetic edge states, where there is additional quantization in terms of "missing flux quanta" \cite{sim1998magnetic}. It would be interesting to explore these phenomena in the context of BDD in future work.


The paper is organized as follows. In Sec. II, we present our mathematical model of the system in interest. This involves setting up the Hamiltonian, solving for the wavefunction, and applying boundary conditions. In Sec. III, we develop an asymptotic approximation to the energy levels of the QD. In Sec. IV, we discussion the results we obtained, including experimental agreement and the validity of our approximation, followed by concluding remarks.


\section{Model}
The QD is modeled as an electron trapped in a cylindrical potential well of radius R and barrier height $V_0$ in a 2D plane. In cylindrical polar coordinates $(r,\phi,z)$, the lateral confinement potential used is given by
\begin{equation}\label{conf_pot}
	V(r) = 
	\begin{cases}
		~0  & r \leq R \\
		~V_{o}  & r > R
	\end{cases}
\end{equation}
In a realistic setting, the QD is also confined along the z-axis due to a cylindrical or lens shape. However, the energy associated with vertical confinement is much larger, and is decoupled from lateral confinement for transition gap measurements in the experiment we are interested in \cite{drexler94}. Note that Equation (\ref{conf_pot}) represents an idealized hard-wall electrostatic confinement, but is better than a parabolic profile as it accounts for the finite lateral size and barrier height of the QD. The QD is placed in a perpendicular magnetic field, which takes a uniform value  $B_i$ inside the QD, and $B_o$ outside the QD respectively.
\begin{equation}
\vec{B}(r) = 
\begin{cases}
~B_{i}~\hat{z} & ~~r \leq R \\
~B_{o}~\hat{z} & ~~r > R
\end{cases}
\end{equation}
Note that we consider an inhomogeneous magnetic field only to keep the analysis general for potential future work. Calculations based on the model presented in Section 4 assume a homogeneous magnetic field: $B_i=B_o=B$. The magnetic field profile corresponds to a continuous magnetic vector potential $\vec{A}(r)$ given by
\begin{equation}\label{vec-pot}
	\vec{A}_p(r) = \left[ \frac{B_{p}r}{2} + \frac{\Phi_{p}}{r} \right]\hat{\phi}, ~~~~{p = i ~\text{or}~ o}
\end{equation}
 The subscript '$p$' can be either '$i$'(inside) or '$o$'(outside), and is helpful in generalizing the analysis inside and outside the QD. In Equation (\ref{vec-pot}), $\Phi_p$ is an intermediate variable with the dimensions of magnetic flux, defined as
\begin{equation}
	\Phi_{p} = 
	\begin{cases}
		~0 & ~~\text{p~=~i} \\
		~\frac{(B_{i}-B_{o})R^{2}}{2} & ~~\text{p~=~o}
	\end{cases}
\end{equation}
The Hamiltonian of the system is that of an electron placed in an electromagnetic field \cite{landau2013quantum}
\begin{equation}
\widehat{H} =  \frac{1}{2m_{p}}\left[\hat{p} + e\vec{A}\right]^{2} + V(r)
\end{equation}
The Hamiltonian commutes with $\partial/\partial\phi$, and hence the wavefunction is separable as $\psi(r,\phi) = e^{il\phi}g(r)$ where $l$ is an integer (0, \textpm 1, \textpm 2 \ldots) The radial part $g(r)$ of the time independent Schrodinger equation $\hat{H}\psi = E\psi$ is hence found to be
\begin{equation}\label{schro}
	-K_{p,l}^2 = \frac{1}{g}\left( g'' + \frac{g'}{r} \right) - \frac{l^2}{r^2} -
	\frac{e^2}{\hbar^2}|\vec{A}(r)|^2 -
	\frac{2le}{r\hbar}|\vec{A}(r)|
\end{equation}
where we define a wave vector $K_p$ as

\begin{equation}
    K_{p,l}^2 = 
	\begin{cases}
		~\dfrac{2m_{i}E}{\hbar^2} & \text{inside}\\
		~\dfrac{2m_o}{\hbar^2}(E-V_o) & \text{outside}
	\end{cases}
\end{equation}
 The solution  of Eq. (6) leads to an exact solution in terms of Kummer functions \cite{ab_steg}. We find, however, that the solution can be well approximated in terms of Bessel functions in a regime defined by constraints on the size and barrier height of the QD,
\begin{equation}\label{st_con}
    R\ll \sqrt{\frac{2\hbar}{e B}} = \sqrt{2}L_m
\end{equation}
\begin{equation}\label{st_con_2}
    V_o\gg \frac{\hbar^2}{2m_oR^2}
\end{equation}
\\Here $L_m=\sqrt{{\hbar}/{e B}}$ is the magnetic length scale and is  also called the Landau length. For $B=1$ T, we require $R\ll 36$ nm. This is reasonable since the radii involved in the experiments of  Drexler \textit{et. al.} was $10$ nm. Further, if $R=10$ nm we require $V_o\gg 5.7$ meV (using $m_o=0.067m_e$), which is a modest lower bound when we use barrier heights of the order of several hundreds of meV or a few eV (around 100 meV in case of the InGaAs-GaAs QD).

Under the regime defined by Eq. (8) and Eq. (9), we find that the radial wavefunction can be approximated as
\begin{equation}\label{bessel}
    g^l_i(r) = A \exp\left( -\frac{e B_i r^2}{4\hbar} \right)J_l(k_i r)
\end{equation}
\begin{equation}
	g^l_o(r) = B \exp\left( -\frac{e B_o r^2}{4\hbar} \right)\frac{\exp(-k_o r)}{\sqrt{r}}
\end{equation}
Here $J_l$ is the $l^{th}$ Bessel function of the first kind. We have also defined wave vectors $k_i$ and $k_o$ as 
\begin{equation}
	k_{i,l}^2 = \frac{2m_iE}{\hbar^2} - (2l+1)\frac{eB_i}{\hbar}
\end{equation}
\begin{equation}
k_{o,l}^2 = \frac{2m_o}{\hbar^2}(V_o-E) + (2l+1)\frac{eB_o}{\hbar} + \frac{e^2R^2}{\hbar^2}B_o(B_i-B_o)
\end{equation}
Note the  last term  in Eq.  (13). If  $B_i=B_o$, the  two expressions
(Eqs. (12)  and (13)) are  the same given the  shift in energy  $E$ to
$V_o-E$.

We now  apply boundary conditions to our  solution (Eq. (10)
and  Eq.(11))   to  obtain  quantized  energy   levels.  Although  the
wavefunction   is    continuous   at   $r=R$,   its    derivative   is
discontinuous. 
\begin{align}
	g^l_i(R) = g^l_o(R)\\
	\frac{dg_i^l}{dr} \bigg|_{r=R} = \beta \ \frac{dg_o^l}{dr} \bigg|_{r=R}  
\end{align}
 Equation (15) is the Ben-Daniel  Duke boundary condition
for our system. As mentioned earlier, $\beta = m_i/m_o$  is the ratio
of effective masses inside and outside the well. 
Eliminating normalization constants, we obtain a non-linear equation for the quantized energy levels of the system.
\begin{equation}\label{nonlin}
	\frac{\beta}{2} ~+~ \beta k_o R ~+~ \frac{eR^2}{2\hbar}(\beta B_o - B_i) ~+~ k_iR \frac{J'_l(k_iR)}{J_l(k_iR)} ~=~ 0
\end{equation}
Equation (\ref{nonlin}) cannot be solved analytically to obtain energy eigenvalues $E$. However, one can obtain a simple asymptotic approximation for the energy levels, which is the subject of our next section.

\section{ASYMPTOTICS}

Consider Eq. (16) for $g_i(r)$, the radial wavefunction inside the QD. For a sufficiently large barrier height, we expect $g_i(r)$ to be close to zero at $r=R$. Equivalently, we expect the argument of $J_l$ to be close to one of its nodes.
\begin{equation}
    k_iR = z_{n l} - \varepsilon
\end{equation}
\\Here $z_{n l}$ is the $n^{th}$ node of $J_l$ and $|\varepsilon|\ll 1$. We can then Taylor approximate $J_l(k_iR)$ as
\begin{equation}
    J_l(k_iR) = J_l(z_{n l} - \varepsilon) \approx -\varepsilon J'_l(z_{n l})
\end{equation}
Hence we have
\begin{equation}
    \frac{J'_l(k_iR)}{J_l(k_iR)} \approx -\frac{1}{\varepsilon}
\end{equation}
Using this result in Eq. (16), we obtain an expression for $\varepsilon$,
\begin{equation}
	\varepsilon = \frac{z_{n l}}{1 + \frac{\beta}{2} + \beta k_oR +\frac{eR^2}{2\hbar}(\beta B_o - B_i) } = \frac{z_{n l}}{\sqrt{\sigma}}
\end{equation}
For large $V_o$ ($\beta k_oR \gg 1$), the largest term in the denominator of Eq. (20) is $\beta k_oR$, where $k_o \approx \sqrt{\frac{2m_oV_o}{\hbar^2}}$. Therefore $\sigma$, as defined in Eq. (20) is approximately given by
\begin{equation}
    \sigma \approx \frac{2m_o}{\hbar^2} (\beta^2 R^2 V_o)
\end{equation}
Using Eq. (17) and Eq. (20) in conjunction with Eq. (12) for $k_i$, we have an asymptotic approximation for the energy levels as
\begin{equation}\label{asym}
	E_{n,l} = \frac{\hbar^2 z_{n l}^2}{2\beta m_oR^2}\left( 1 - \frac{1}{\sqrt{\sigma}} \right)^2 + \left( l+\frac{1}{2} \right)\frac{\hbar e B_i}{\beta m_o} 
\end{equation}
Equation (\ref{asym}) has an elegant physical meaning. Suppose we had an electron trapped in cylindrical potential well with radius $(R+\delta)$, where $\delta=\frac{R}{\sqrt{\sigma}}$, without an external magnetic field. The first term of Eq. (\ref{asym}) represents quantized energy levels of the aforementioned system. If we now switch on a perpendicular magnetic field $B_i$ inside the well, additional Landau levels are observed whose splitting energy is described by the second term of Eq. (\ref{asym}). $\delta$ can thus be interpreted as a penetration depth of the wavefunction due to lateral confinement. 


The levels are hence classified by quantum numbers $(n,l)$. The ground
state  of the  QD is  (1,0), while  the next  five states  are (1,-1),
(1,1),  (1,-2), (1,2)  and (2,0).   The series  is generated  from the
relative  locations of  $z_{n l}$,  the  $n^{th}$ root  of the  Bessel
$J_l(x)$,      which      displays      the      following      trend:
$z_{10}<z_{11}<z_{12}<z_{20}$.   In the  presence  of a  perpendicular
magnetic field, the states $(n,l)$ and $(n,-l)$ lose their degeneracy,
resulting  in  Landau   level  splitting  with  a  gap   $\Delta  E  =
2l\hbar\omega_i$  where   $\omega_i=eB_i/m_i$.   We  also   expect  an
additional  but  smaller, $g\mu_B  B$  splitting  between spin-up  and
spin-down  electron states,  well known  as Zeeman splitting.  We
however ignore Zeeman  splitting, and only consider the effect of BDD on spin-degenerate Landau levels.

For $V_0 = 100$ meV, $m_o=0.067m_e$, $\beta=0.7$, $R=11$ nm, homogeneous magnetic field $B_i=B_o=B$, and considering the states $n=1$, $l=\pm1$, the asymptotic approximation reads
\begin{equation}
    E_{1,l=\pm1} = \left[60.16\left( 1-\frac{B}{717} \right)^2 + \left( l+\frac{1}{2} \right)2.47 B\right] ~meV
\end{equation}

where $B$ is in Tesla. Notice that the quadratic dependence on $B$ is negligibly small for the range of magnetic field we are interested in and even beyond. This is why we do not see a curvature in the Landau levels (Figure 3) even for high magnetic fields. For $R=10$ nm, the Bessel approximation (Equation (\ref{bessel})) is applicable only when $B\ll 13$T. We expect the Bessel approximation to gradually break down for $B>5$T. For higher magnetic fields, the experimentally expected curvature in the Landau levels can only be observed by deriving the energy eigenvalues using the general solution to Equation (\ref{schro}) in terms of Kummer functions, which are notorious to deal with. For the purpose of this work, we limit ourselves to the effect of BDD on these approximately linear levels.

\section{RESULTS}
\subsection{CHARGE DENSITY PROFILE}
The  radial  charge  density of  the  QD  is  given  by $\widetilde{\rho}(r)  =  e 2\pi r
g^2(r)$ so that $\int \widetilde{\rho}(r)dr = e$.  We investigated  the scaled charge  density  profile $\rho(r) = rg^2(r)$ of  a QD  with
radius 10 nm, confined  with $V_o$ = 1 eV, and  under a magnetic field
of 1 T.

In Fig. 1, we examine the effect of $\beta$ on the ground state charge
density  profile. We  observe that  charge spreads  out closer  to the
boundary as  $\beta$ is decreased.  Further, a small amount  of charge
leaks out of  the QD for small $\beta$ and this  is enhanced as
$\beta$ is decreased.  This is an interesting observation,  and can be
explained by the fact that  the tunnelling probability at the boundary
of  the QD  decays exponentially with the  difference $(V_o-E)$.  As
$\beta$ decreases,  energy levels rise and the tunnelling probability increases. One  can also  see the discontinuity  in the  derivative of
$\rho(r)$  at  the boundary  for  small  $\beta$, which  results  from
imposing BDD.

The inset in Fig. 1 shows  the charge density profile of states (1,0),
(1,1) and (1,2). The states $(n,l)$  and $(n,-l)$ have the same charge
density since  they only  differ by an overall  phase ($e^{i2l\phi}$).  We note
that higher energy levels are associated with more spreading of charge
towards the boundary. The peak charge density is larger, and is closer
to the boundary as we consider higher levels.

\subsection{DEPENDENCE OF TRANSITION ENERGIES ON  SIZE}
Energy difference  is the object  of study in absorption  and emission
spectra. Hence, in Fig. (2) we investigate the effect of size, $\beta$,
and  the magnetic  field on  the transition  energies $(E_1-E_0)$  and
$(E_2-E_1)$  (Fig.  2).   We observe  that $(E_1-E_0)$  decreases with
increasing  size,  while  $(E_2-E_1)$  has  almost  no  dependence  on
size. We fit $(E_1-E_0)\propto  1/R^\gamma $ using Levenberg-Marquardt
fit. The  values of the  exponent are found  to be $\gamma  =1.76$ for
$\beta=0.1$ and  $\gamma = 2.11$  for $\beta=1$.  This lateral size confinement effect be explained  well  by the  asymptotic expression  we
developed in  Sec. III.  From Eq.  (22), we  note that  the transition
energies of interest are determined to be
\begin{equation}
    (E_1-E_0) \approx \frac{8.9\hbar^2}{2\beta m_oR^2}\left( 1 - \frac{1}{\sqrt{\sigma}} \right)^2 - \frac{\hbar e B_i}{\beta m_o}
\end{equation}
\begin{equation}
    (E_2-E_1) \approx \frac{2\hbar e B_i}{\beta m_o}
\end{equation}
From Eq.  (23)  and Eq.  (24), we note that  $(E_1-E_0)$ has a $1/R^2$
dependence on size,  while $(E_2-E_1)$ has no dependence  on size.  We
also examined the effect of  $\beta$ on these transition energies. The
transition  energies have  a  strong dependence  on  the magnitude  of
$\beta$. As  evident from  Fig. 2,  both transition  energies increase
sharply as $\beta$  is decreased. This is expected  from the $1/\beta$
dependence  predicted by  Eq.  (23) and  Eq.  (24). Transition  energy
$E_2-E_1$ increases  two times  when magnetic  field is  increased ten
times. However, the same transition energy has stronger dependence on
BDD condition. Energy $(E_2-E_1)$ is increased seven times on decreasing
$\beta$ ten times. 

\subsection{DEPENDENCE OF ENERGY LEVELS ON MAGNETIC FIELD}
Figure 3 shows  the energy levels of  our system as a  function of the
applied magnetic field.
Interestingly the plots are  linear in
$B$. This can be understood on the basis of our asymptotic
analysis (Eq. (22)). Cyclotron energy ($\hbar \omega_i/2$) is manifestly
linear in $B$. Additionally a detailed analysis revealed that the
first term in Eq. (22)  ($1/{\sqrt{\sigma}}$) is also
linear in $B$. Thus the exact results also indicate a linear trend of energy on magnetic field.

We observe clear Landau level splitting as the
magnetic  field is  increased. The  effect  of imposing  BDD is  again
evident from  the sharp fall  in the  energy levels for  $\beta=1$. We
also   find  further   support   for   our  asymptotic   approximation
(represented by dashed lines), which converges well with our numerical
results  for  large  magnetic  field  strengths.   Asymptotic  equation
(Eqs.  (20) and  (22))  points  out an  important  fact. Changing  the magnetic  field outside does  not radically affect  the energy  of the QD. This can be explained by  the fact that probability of finding an
electron in  the outer  region of  the dot  is very  low. If  \(B_{o} =
{B_{i}}/{\beta}\)    the    effect    of    magnetic    field    in
$\sqrt{\sigma}$  is zero . Which  brings out the fact that we can
use $B_{o}$  as factor  to make $B$  affect $E$ only  in terms  of the
Landau energies.  We carried a fit to energy of the form
 \begin{equation} 
E = \frac{C}{R^\gamma}
 \end{equation}
Table (\ref{tab:fit_EvsR}) lists the values of the
 exponent $\gamma$. Interestingly we find that BDD effect reduces 
$\gamma$. The  magnetic field also tends to reduce $\gamma$ but
marginally so. This can perhaps be understood as follows: increasing
$B$ reduces the cyclotron radius of the electron and hence the electron density at the dot boundary is inconsequential.
{%
\newcommand{\mc}[3]{\multicolumn{#1}{#2}{#3}}
\begin{table}
\begin{center}
\begin{tabular}{|l|l|l|}\hline\hline
\mc{1}{|c|}{$B$ (T)} & \mc{1}{|c|}{$\beta$} & $\gamma$\\\hline\hline
\mc{1}{|c|}{0} & \mc{1}{l|}{0.1} & \mc{1}{l|}{1.46}\\\cline{2-3}
\mc{1}{|l|}{} & \mc{1}{l|}{1.0} & \mc{1}{l|}{1.91}\\\hline
\mc{1}{|c|}{1} & \mc{1}{l|}{0.1} & \mc{1}{l|}{1.56}\\\cline{2-3}
\mc{1}{|c|}{} & \mc{1}{l|}{1.0} & \mc{1}{l|}{1.91}\\\hline
\mc{1}{|c|}{10} & \mc{1}{l|}{0.1} & \mc{1}{l|}{1.22}\\\cline{2-3}
\mc{1}{|l|}{} & \mc{1}{l|}{1.0} & \mc{1}{l|}{1.81}\\\hline
\end{tabular}
\caption[QD in transverse magnetic field: investigating the dependence
  of $E$ on $R$ as 
  $ E = C / R^\gamma$ for a 2DQD.]{Investigating the dependence of $E$ on $R$ as
  $ E = C / R^\gamma$. Table gives the values of $\gamma$ for
  various values of $\beta$ and $B$. }  \label{tab:fit_EvsR} 
\end{center}
\end{table} 
}

\subsection{COMPARISON WITH EXPERIMENT}

Drexler \textit{et. al.} \cite{drexler94}, in their experiment, used an In\textsubscript{0.5}Ga\textsubscript{0.5}As-GaAs QD ($m_i=0.047m_e$ and $m_o=0.067m_e$, so $\beta\approx0.7$) with  a radius of
$(10\pm 1)$  nm. They measured  the transition gap $(E_2-E_0)$  as a function of an  applied perpendicular magnetic field,  and found their data to agree with a parabolic confinement model using $m^{*}=0.07m_o$ and $\hbar\omega_o  = 41$  meV. Also, their dots are lens-shaped along the z-axis. However, they mention that their measured transition gap data from IR spectroscopy is decoupled from the high energy z-confinement, and is only due to lateral confinement. The bandgap difference between In\textsubscript{0.5}Ga\textsubscript{0.5}As and GaAs is about $360$ meV. But this is further split between valence and conduction band offsets, so a realistic value of the barrier height is closer to $100$ meV. In the present  work, we have developed a single-electron hard-wall confinement model including BDD within the effective mass framework. It is also important to note that our model is best applicable in a strong confinement regime described by Equations (\ref{st_con}) and (\ref{st_con_2}). Using $V_0 = 100$ meV and $\beta=0.7$,  we find good agreement  with their experimental data  for a QD radius  of $11$ nm (Figure 4).

In conclusion, we presented a hard-wall confinement based model for a 
circular Quantum  Dot placed in  a perpendicular magnetic  field. Most
importantly,  we  demonstrated  the  strong  influence  of  using  the
BenDaniel-Duke  boundary  condition  on   the  set of spin-degenerate Landau levels  of  the
system. We also  developed a simple asymptotic  approximation (Eq. (\ref{asym})) to the energy levels, which is in fair agreement with numerical results. The agreement is enhanced  for larger size and confinement potential. Finally, we  also observed  that a  particular transition gap in  our model agrees     well     with      experiments     performed     on     the
In\textsubscript{0.5}Ga\textsubscript{0.5}As-GaAs      Quantum     Dot
\cite{drexler94}.

\section{ACKNOWLEDGEMENTS}
We are extremely thankful to our mentor Praveen Pathak (HBCSE-TIFR) for regular discussions and guidance. We also also thank Vijay Singh (HBCSE-TIFR) for useful discussions. We acknowledge the support of the Govt. Of India, Department of Atomic Energy, under the National Initiative on Undergraduate Science (NIUS) of HBCSE-TIFR (Project No. 12-R\&D-TFR-6.04-0600).
\printcredits

\newpage

\begin{figure}
    \centering
    \includegraphics{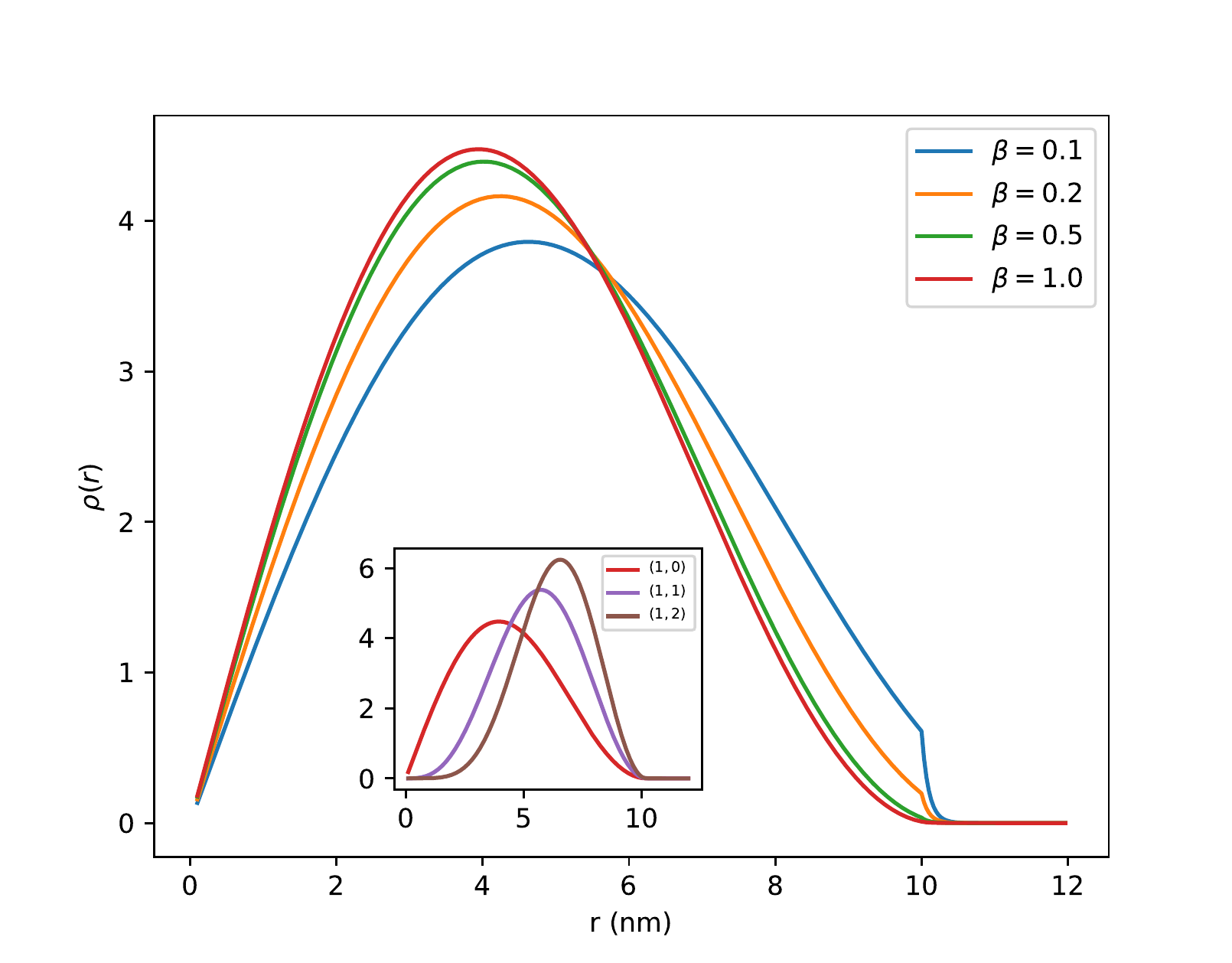}
    \caption{Scaled radial charge density ($rg^2(r)$) profile of a QD with R = 10 nm, B = 1 T, and $V_o$ = 1 eV. The main plot shows the effect of $\beta$ on the ground state charge density profile. The inset shows the charge density profile of levels (1,0), (1,1) and (1,2) for $\beta$=1. ($m_o=m_e$, $m_i=\beta m_o$)}
    \label{charge_dens}
\end{figure}

\begin{figure}
    \centering
    \includegraphics{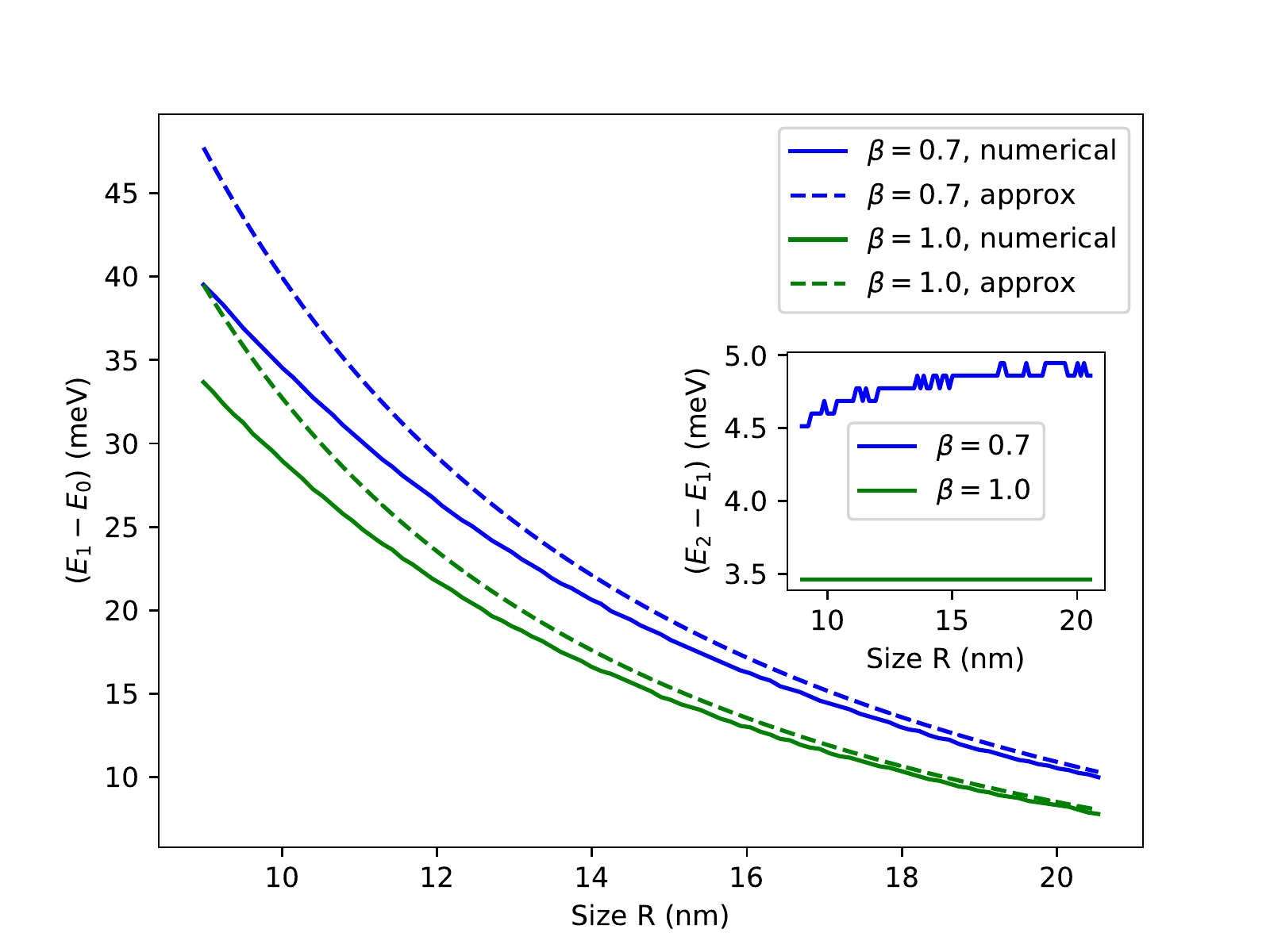}
    \caption{Transition energies as a function of the radius of the QD
      ($m_o$ = 0.067$m_e$, $V_o$ = 100 meV, B = 1 T) in  the range of 10 nm to 20 nm. The main plot  shows the  effect of  size and  $\beta$ on  the transition gap $(E_1-E_0)$. The inset shows the same for the transition gap  $(E_2-E_1)$. The dashed lines in the main plot represent the asymptotic approximation (Eq. (22)). The case of $\beta = 0.7$ corresponds to the InGaAs-GaAs QD ($m_i$ = 0.047$m_e$, $m_o$ = 0.067$m_e$). }
    \label{trans_energy}
\end{figure}

\begin{figure}
    \centering
    \includegraphics{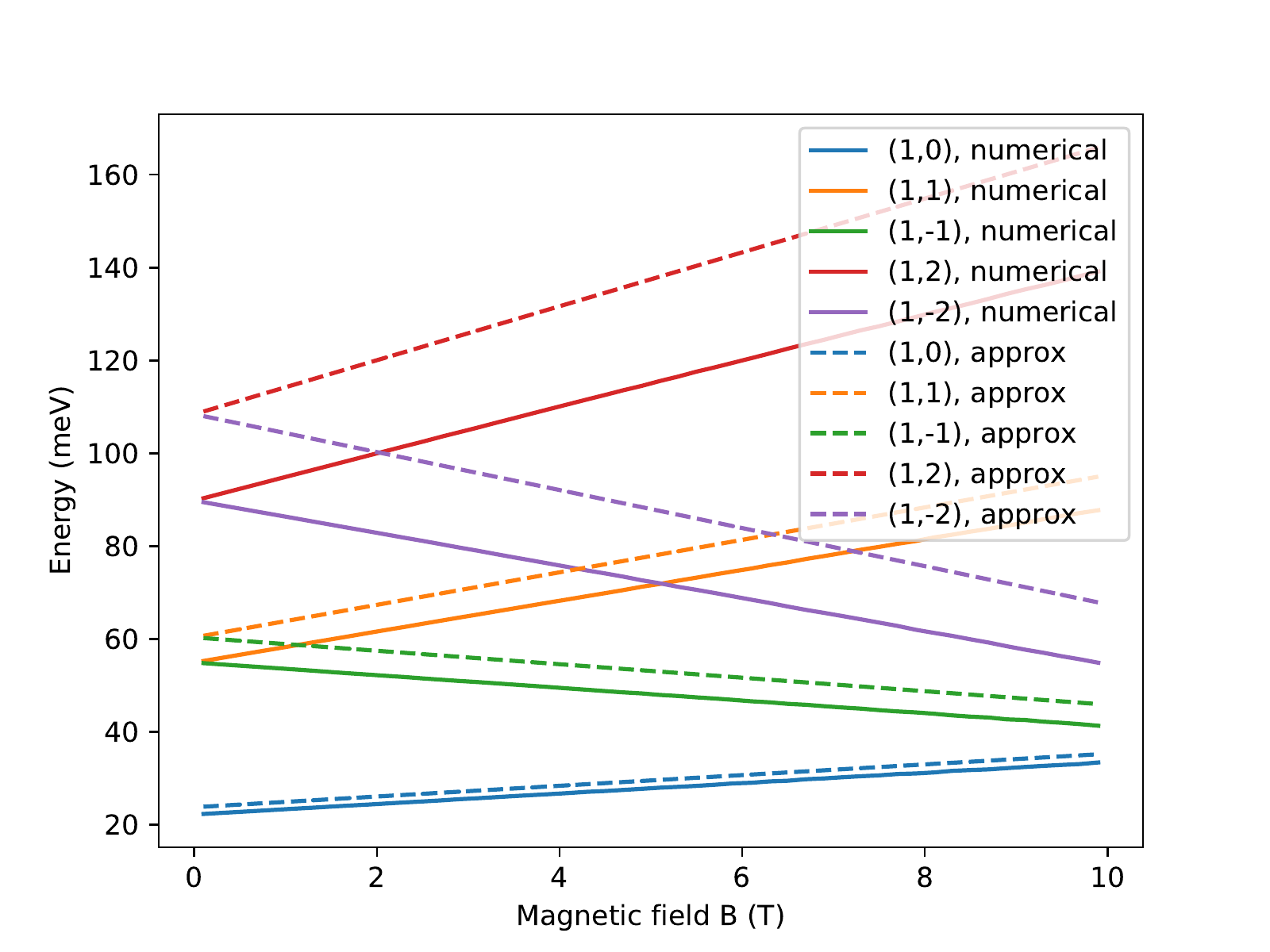}
    \caption{Energy levels of the InGaAs-GaAs QD (R = 11 nm, $V_o$ = 100 meV, $m_o$ = 0.067$m_e$, $\beta$ = 0.7) as a function of an applied magnetic field. The solid lines depict the lowest five levels as obtained from numerical computation. The dashed lines depict the same five levels as expected from our asymptotic approximation (Eq. (22))}
    \label{E_vs_B}
\end{figure}

\begin{figure}\label{exp_fit}
    \centering
    \includegraphics{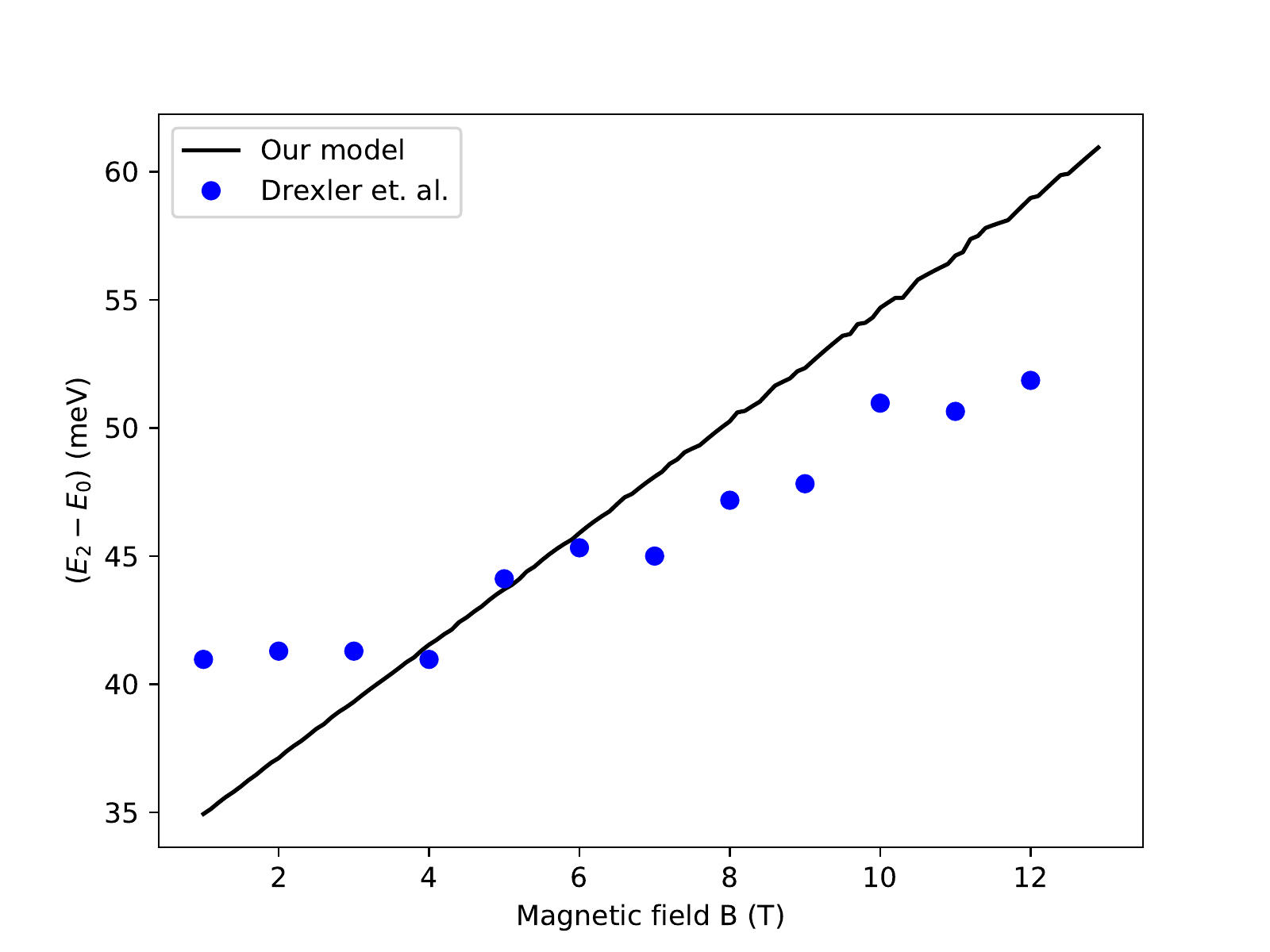}
    \caption{The full line (black) depicts transition energy $(E_2-E_0)$ of the QD ($V_o$ = 100 meV, $\beta$ = 0.7, R = 11 nm) as a function of the applied magnetic field. The blue circles are experimental data \cite{drexler94} on the same transition energy for a GaAs-InGaAs QD with R = (10\textpm 1) nm. ($m_i = 0.047m_e$, $m_o = 0.067m_e$)}
    \label{expt_fit}
\end{figure}

\end{document}